\documentstyle[prl,aps,multicol,epsf]{revtex}
\begin{document}

\draft
\title{Resonant transmission through an open quantum dot}
\author{C.--T.~Liang, I.~M.~Castleton, J.~E.~F.~Frost, C.~H.~W.~Barnes,
C.~G.~Smith, C.~J.~B.~Ford, D.~A.~Ritchie, and M.~Pepper
                                                                       }

\address{
Cavendish Laboratory,
Madingley Road,
Cambridge CB3 0HE,
United Kingdom
}

\date{\today}

\maketitle

\widetext
\begin{abstract}
\leftskip 54.8pt
\rightskip 54.8pt
We have measured the low-temperature transport properties of a quantum dot
formed in a one-dimensional channel.
In zero magnetic field this device shows
quantized ballistic conductance plateaus with 
resonant tunneling peaks in each transition region between plateaus.
Studies of this structure as a function of applied
perpendicular magnetic field and source-drain bias indicate that
resonant structure deriving from tightly bound states is split by Coulomb
charging at zero magnetic field.
\end{abstract}

\begin{multicols}{2}
\narrowtext
 
Advancing
technology has made it possible to define artificial semiconductor
microstructures which confine electrons in all three spatial
dimensions\cite{CGSdot} with discrete zero-dimensional states. Such
structures, often called quantum dots, provide uniquely simple systems
for the study of few electron physics. In particular, the Coulomb
blockade (CB) of single electron tunneling through quantum
dots\cite{Meirav}
has been extensively investigated \cite{CB}. It has been  
demonstrated \cite{McEuen91} that transport through small quantum dots
is determined
by charging effects\cite{Ford,Weis} as well as quantum confinement effects 
\cite{Su,Ash,Johnson}.
Quantum dots can also be formed by impurities which are either
directly in the electron gas, as for Si devices \cite{Scott}, or are 
remote ionized donors in a spacer
layer\cite{Davies89} as for the GaAs/Al$_{x}$Ga$_{1-x}$As
heterojunction \cite{JTN}. The CB effects in
such unintentionally defined quantum dots have been studied extensively
\cite{Scott,JTN,Pepper}.

Within a non-interacting picture Tekman and Ciraci\cite{Tek} have 
predicted that
resonant tunneling (RT) may occur through energy states bound to an attractive
impurity potential in a split-gate device 
even when some one-dimensional (1D) channels are perfectly 
transmitted. Therefore in
addition to 1D quantized conductance steps \cite{vanWees,Wharam}, replicated
resonant features between plateaus should be observed when
a quantum dot formed by an impurity potential is present in a
split-gate device.
In this paper, we report the first observation of such resonant structure
from a quantum dot formed by an impurity potential in a split-gate
device.  We show how these RT features develop in a
perpendicular magnetic field $B$ and we investigate the energy spacings between
different resonant states using source-drain bias measurements. 
   
The Schottky gate pattern shown in the inset to Fig.~1
was defined by electron beam lithography on the surface of a
GaAs/Al$_{0.3}$Ga$_{0.7}$As heterostructure, 90 nm above a two-dimensional
electron gas (2DEG). The carrier concentration of the 2DEG was
$3.3\times 10^{15}$ m$^{-2}$ with
a mobility of 90 m$^{2}$/Vs.
Experiments were performed in a dilution refrigerator at 100 mK and
the two-terminal differential conductance $G=dI/dV$ was measured using
an ac excitation voltage of 10 $\mu$V.
 
Figure~1 shows the differential conductance as a function
of the voltage $V_{g1}$ on gate~1, for various 
voltages $V_{g2}$ on gate~2. For $V_{g2}$=--1.7 V (trace~3)
we observe replicated resonant 
peaks in $G(V_{g1})$, reminiscent of those predicted\cite{Tek}.
As the temperature was
increased, these structures became broader but were
still discernible up to 650 mK.
When the conduction channel through the
split-gate structure was moved sideways by varying the 
voltage on gate 2 \cite{Rudi} away from $V_{g2}$=--1.7 V, the sharp RT 
features gradually diminished until at $V_{g2}$=--2.6 V (trace~7)
only quantized 
1D ballistic conductance steps were seen. In a
subsequent cooldown in a $^3$He cryostat, we did not observe 
identical RT structure. Although the surface Schottky gate pattern was
intended to define a quantum dot in the 2DEG electrostatically,  
both observations suggest that
ionized impurities in the spacer layer \cite{Davies89} played an
important role in
determining the transport properties through the channel defined by
the surface gate. Since we observe conductance peaks
(resonant tunneling) rather than resistance peaks (resonant reflection),
we believe that in our system an
attractive impurity potential helped create a quantum dot.
Previously McEuen {\em et al.\/}\cite{McEuen90} claimed that
two resonant transmission peaks they observed for $G < 2e^2/h$ 
in a disordered split-gate device derived from  
the formation of a quantum dot by a {\em single\/} hydrogenic impurity.
In this experiment only two peaks were observed because the electrons
which filled the impurity bound states acted to screen it so that at higher
energies only quantized conductance with no resonant structure was seen.
In our experiment we observe at least fourteen resonant peaks (see trace 3
in Fig. 1), implying that the impurity potential does not become screened
even after accommodating 14 electrons. We do not believe that such a 
potential could be generated by a single ionized impurity, only a cluster
would be capable of this.

Figure 2 shows $G(V_{g1})$ for $V_{g2}$=--1.7 V at
different $B$. 
For $G < 2e^2/h$, the two RT peaks have a weak
$B$ dependence and persist to $B= 4$~T. As the magnetic field  
is increased, the conductance plateaus and the
RT peak positions for $2e^2/h < G < 8e^2/h$ move to
more positive $V_{g1}$ as a result of the formation of hybrid
magnetoelectric subbands\cite{Ash}. At $B \approx 2$~T these features 
are no longer seen. A broad RT 
peak adjacent to the sharp RT peaks for $4e^2/h < G < 6e^2/h$ develops
at $B=0.6$~T and splits into two at higher magnetic fields, as
indicated by arrows. Similar but less pronounced results can be also
seen for $2e^2/h < G <4e^2/h$.

The sharp resonances correspond to tightly bound states and the 
broad resonances to weakly bound states within the picture of Tekman 
and Ciraci\cite{Tek}.
The application of a perpendicular magnetic field strengthens the 
confinement of states in a quantum dot by localizing electron wavefunctions 
to the sample boundaries \cite{mace1}. This is consistent with the
disappearance of the tightly bound states (they become immeasurably small) 
and the strengthening of the resonant structure from the weakly bound states
at high field seen experimentally - Fig.~2.
When the sharp RT 
structures for $2e^2/h< G < 6e^2/h$ have disappeared,  
oscillations in $G(V_{g1})$ are still observed.  
Their structure is more complicated and possibly derives from 
a combination of resonant transmission and resonant reflection 
\cite{JK} from bound states.

We now discuss the separation in gate voltage $\Delta V_{g1}$ between each pair
of tightly bound RT peaks at various magnetic fields (marked as square,
circle, triangle and cross in Fig~2).
Figure~3 shows $\Delta V_{g1}(B)$ for RT features which occur with
different numbers of transmitted 1D channels $n_{1D}$.
For $n_{1D}$=0, $\Delta V_{g1}$ shows only a weak magnetic field
dependence. 
For $n_{1D}$=1, 2, and 3, $\Delta V_{g1}$ shows saturation at low $B$ and 
a linear $B$ dependence at high $B$.

At $B=0$, $\Delta V_{g1}$
decreases as $n_{1D}$ increases, as shown in the inset to Fig.~4.
To obtain the energy spacing $\Delta E(n_{1D})$
between pairs of tightly bound RT peaks, we have used a standard 
source-drain bias technique
\cite{Weis,Johnson,JTN,Nalin}. 
$\Delta E$ decreases
dramatically from $n_{1D}$=0 to $n_{1D}$=1 (see Fig.~4). Note
that we were not able to measure $\Delta E$
between pairs of RT peaks for $G > 8e^2/h$,
perhaps because the application of a dc bias caused the quantum dot to
break down.

Within the non-interacting picture
\cite{Tek} at $B=0$ the energy states through
which RT occurs are spin-degenerate.
As $B$ is increased, if there is no spin-splitting, states with 
different angular momentum in the same Landau level become closer in 
energy\cite{Darwin}. If the Zeeman energy is included,
electrons in the same Landau level with the same angular momentum
but different spin move apart in energy
causing individual resonant transmission peaks to split into two peaks.
Using the minimum possible value 0.44 for 
the Land\'{e} g-factor in our system, we
estimate the Zeeman energy to be $\approx0.1$~meV at $B=4$~T,
a factor of twelve
larger than thermal smearing at 100~mK, and equal the
full-width-half-maximum of the tightly bound peak
closest to pinch-off, suggesting that such
splitting would be observable in our system. However, as shown in Fig.~2,
the individual peaks in each pair of tightly bound RT peaks 
{\em do not\/} split at any magnetic field. In addition pairs of peaks
{\em do not\/} come closer
together and each pair of peaks remains in the same transition region.
These factors imply that each pair of peaks derives from the same 
single-particle state. They split at zero magnetic field due to
the energy difference between 
single and double occupation of a single state \cite{Ashmin}.
However, the case of charging-induced splitting in mesoscopic devices 
where two adjacent single-electron tunneling peaks are related
to states with different spin quantum numbers \cite{Weis2}, 
is only well understood in the Coulomb blockade regime \cite{Meir}
for $G < 2e^2/h$. Assuming that the relations
$\Delta E=30.1\Delta V_{g1}$~meV/V ($n_{1D}=0$),
$\Delta E=14.8\Delta V_{g1}$~meV/V ($n_{1D}=1$),
$\Delta E=10.3\Delta V_{g1}$~meV/V ($n_{1D}=2$), and
$\Delta E=7.27\Delta V_{g1}$~meV/V ($n_{1D}=3$) (determined from
the data shown in Fig.~4 and the inset) which hold at
$B=0$ are still valid at high field,
$\Delta E(B)$ for the tightly bound peaks with $n_{1D}$=1, 2,
and 3 shown in Fig.~5 also implies charging-induced 
splitting at $B=0$. If the splitting arose solely
from Zeeman splitting, then one would expect $\Delta E(B)
\rightarrow 0$ as $B \rightarrow 0$.
Instead $\Delta E(B)$ shows saturation at low $B$, suggesting that the
splitting at low fields is due to some effect other than
Zeeman splitting. The linear fits 
$\Delta E=0.636B$ (solid line), $\Delta E=0.507B$ (dotted line), and
$\Delta E=0.424B$ (dashed line)
shown in Fig.~5 yield Land\'{e}
g-factors of 10.9, 8.7, and 6.9 for $n_{1D}=1, 2$, and 3, respectively.
Such large g-factors have been measured in the quantum Hall regime
where exchange energy is important \cite{Nicholas}. 
For the case $n_{1D}$=0, $\Delta E(B)$ has a weak
$B$ dependence since near pinch-off the Coulomb charging effect is
much stronger than the Zeeman term.
  
Having established the role of Coulomb charging effects in our system,
we can now explain the splitting of the broad resonant tunneling
peak, indicated by arrows in Fig.~2, which occurs as $B$ is increased. For
$B=0.6$~T, the broad resonant tunneling peak is spin-degenerate, as
the state through which RT occurs is weakly bound and the Coulomb
charging arising from confinement is not pronounced. At higher $B$
this state becomes more tightly bound, increasing the
contribution of Coulomb charging effects. Therefore when the applied magnetic
field is increased from $B=0.6$~T, both the Zeeman term
and the Coulomb charging lift the electron spin-degeneracy, causing
the broad resonant tunneling peak to split into two.

The decrease of $\Delta E(n_{1D})$ as $n_{1D}$ is increased,
at $B=0$, shown in Fig.~4, arises from two
mechanisms:  the Coulomb force between electrons bound in the quantum dot
is increasingly screened as $n_{1D}$ is increased; and the
conduction channel defined by the surface Schottky gates becomes
wider, increasing the spatial extent of the bound state wavefunctions,
and hence reducing the Coulomb charging energy as $n_{1D}$ is increased. 

Although we can explain our results in terms of Coulomb
charging effects qualitatively, it is important to note that 
ascribing the pairs of sharp RT features 
to zero-field splitting, for  $G > 2e^2/h$,
requires an extension of the Coulomb charging picture to
the metallic regime  where some 1D channels are transmitted, and that the
Coulomb interaction between pairs of electrons are partially screened
by these 1D channels.
In principle the results we present here are able to
give information on the ability of 1D states to screen 0D states.

In conclusion, we have reported an observation of
transmission resonances through an open quantum dot. 
The magnetic field dependence of pairs of tunneling peaks 
provides experimental evidence for Coulomb charging effects at
zero-field magnetic field
even when some one-dimensional channels are perfectly 
transmitted through the open quantum dot. 

This work was funded by the United Kingdom
(UK) Engineering and Physical Sciences Research Council. We thank 
D.H.~Cobden, J.H.~Davies, D.E.~Khmelnitskii, P.C. Main, 
S.E.~Ulloa, and in particular J.T. Nicholls  
for helpful discussions. C.T.L. acknowledges financial support from
Hughes Hall College, the Committee of Vice--Chancellors and Principals, 
UK, and the C.R. Barber Trust Fund. C.H.W.B.
acknowledges support from the Isaac Newton Trust.

\centerline{\bf Figure Captions}

Figure 1.
$G(V_{g1})$ when the conduction path is
electrostatically shifted by applying various gate voltages to gate 2.
Trace 1 to 7: $V_{g2}$= $-1.3$, $-1.5$, $-1,7$, $-1.9$, $-2.1$,
$-2.3$, and $-2.6$~V, respectively. The inset shows the Schottky gate
geometry. 

Figure 2.
$G(V_{g1})$ for $V_{g2}$=$-1.7$ V
at various magnetic fields. The corresponding magnetic fields are,
from bottom to top: $2.5$, $2.4$, $2.3$, $2.2$, $2.1$, $2$, $1.9$,
$1.8$, $1.7$, $1.6$, $1.5$, $1.4$, $1.2$, $1$, $0.8$, $0.6$, $0.4$,
$0.2$, and 0 T. Traces are vertically offset for
clarity. The symbols
indicate the evolution of the RT features for $G < 2e^2/h$ (square),
$2e^2/h < G <4e^2/h$ (circle),
$4e^2/h < G < 6e^2/h$ (triangle), and $6e^2/h < G < 8e^2/h$ (cross)
as the applied
magnetic field is increased from 0~T. Arrows serve as a guide to the
eye indicating a single resonant  peak splits into two for 
$2e^2/h < G < 4e^2/h$ and $4e^2/h < G < 6e^2/h$, respectively.

Figure 3.
$\Delta V_{g1}(B)$ for various $n_{1D}$.

Figure 4.
The energy spacing $\Delta E$ between 
pairs of RT peaks as a function of $n_{1D}$ deduced from the dc bias 
measurements. The inset shows $\Delta V_{g1}(n_{1D})$.

Figure 5.
The energy spacing $\Delta E$ between
pairs of RT peaks as a function of $B$ determined from data shown in
Fig.~3 and 4. The straight line fits are discussed in the text.

\end{multicols}
\end{document}